\documentclass[sigconf]{acmart}
\usepackage{natbib}
\usepackage[linesnumbered,ruled,vlined]{algorithm2e}
\usepackage{listings}
\usepackage{hyperref}
\usepackage{graphicx}
\usepackage{bbding}
\usepackage{float}
\usepackage{xcolor}
\AtBeginDocument{%
	\providecommand\BibTeX{{%
			\normalfont B\kern-0.5em{\scshape i\kern-0.25em b}\kern-0.8em\TeX}}}

\begin{document}
	
	\title{Improving API Documentation Comprehensibility via  Continuous Optimization and Multilingual SDK}
	\author{Shujun Wang}
	\email{wangshujun.wsj@alibaba-inc.com}
	\affiliation{%
		\institution{Alibaba Group}
		\city{Beijing}
		\country{China}
		\postcode{100000}
	}
	
	\author{Yongqiang Tian}
	\email{puling.tyq@taobao.com}
	\affiliation{%
		\institution{Alibaba Group}
		\city{Beijing}
		\country{China}
		\postcode{100000}
	}
	
\author{Dengcheng He}
\email{dengcheng.hedc@alibaba-inc.com}
\affiliation{%
	\institution{Alibaba Group}
	\city{HangZhou}
	\country{China}
	\postcode{100000}
}

	\begin{abstract}
		Optimizing and maintaining up-to-date API documentation is a challenging problem for evolving OpenAPIs. In this poster, we propose a data-driven continuous optimization solution and multilingual SDK generation scheme to improve the comprehensibility of API documentation. We compute the correlation between API integrity and API trial success rate. Based on this, we partition the API to ensure that each API has a correct optimization direction. Then, we propose a fine-grained(i.e., parameter level) continuous optimization solution to annotate problems in API documents in real-time. Based on the above resolutions, we can provide theoretical analysis and support for the optimization and management of API documents. Finally, we explore the crucial challenges of OpenAPIs and introduce a tailored solution, TeaDSL, a multi-language SDK solution for all OpenAPI gateways. TeaDSL is a domain-specific language that expresses OpenAPI gateways, generating SDKs, code samples, and test cases. The experiments evaluated on the online system show that this work's approach significantly improves the user experience of learning OpenAPIs.
	\end{abstract}
	
	\keywords{API,
		Data-driven,
		Optimization,
		Real-time}
	
	\maketitle
	
	\section{Introduction}
	
	Application Programming Interfaces (APIs) play an essential role in modern software development\cite{API}. OpenAPI indicates a behavior where producers offer Application Programming Interfaces (APIs) to help end-users access their data, resources, and services. With the help of OpenAPIs, developers can complete their tasks more efficiently\cite{APIEffecient}. Thus, OpenAPI has received extensive attention from industry and academia\cite{apiuse1,apiuse10,apiuse11,apiuse12}. Table~\ref{tab:doc} exhibits a fragment of API documentation. Researchers identified the documentation of APIs as the primary source of information and the critical obstacle to API usability. In this regard, researchers have identified the qualities of "good API documentation" as follows: complete, correct, includes thorough explanations and code examples, and provides consistent presentation and organization.  
	\begin{table}
		\caption{A Fragment of API Documenration\label{tab:doc}}
		\begin{tabular}{|c|c|c|c|c|} 
			\hline
			Parameter & Type & Required & Description & Example\\
			\hline 
			\scriptsize TemplateCode & \scriptsize String & \scriptsize Yes & \scriptsize Message Template ID&\scriptsize SMS\_123456 \\ 
			\hline
            \scriptsize PhoneNumber & \scriptsize String & \scriptsize Yes & \scriptsize Phone Number &\scriptsize 186****9602 \\
			\hline
		\end{tabular}
	\end{table}
	
	As of January 2023, Alibaba Cloud lists over 14500 OpenAPIs for web services(https://next.api.aliyun.com/home), and these APIs currently serve hundreds of thousands of users. Based on an ocean of users interacting with the API documentation, we detected that the "Comprehensibility" of the documentation should also be a component of "good documentation." Actually, "Comprehensibility" is an empirical challenge for end-users since the knowledge gap between API designers and API users\cite{gapbetuseranddesigner}.
	We continue to optimize the API documentation based on user feedback to mind this gap. However, the feedback-based document optimization solution has two significant shortcomings: 
	The amount of feedback within a period is limited, and user feedback is usually coarse-grained, such as "document problem"  and "document unclear." 

\begin{figure}[htbp]
		\centering
		\includegraphics[width=0.4\linewidth]{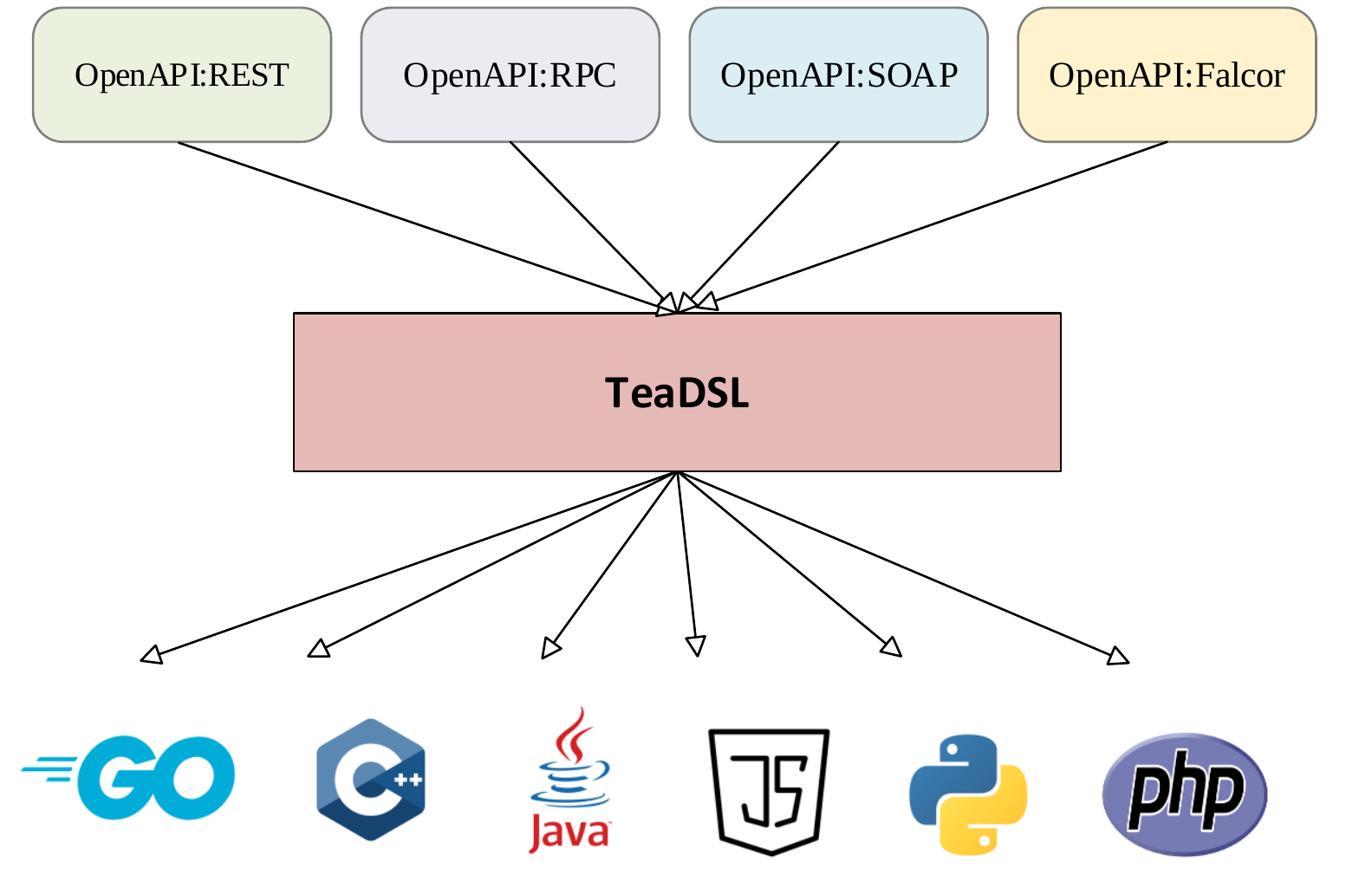}
		\caption{TeaDSL SDK Generation}
		\label{TeaDSL}
	\end{figure}

	Besides, many researchers have pointed out that multilingual SDK code samples are essential to API documentation. However,  existing work for SDK generation mainly focuses on generating isolated SDKs. In this case, the SDK and OpenAPI gateway are highly coupled, which means that any change in the OpenAPI gateway will cause all SDKs to be rewritten. 
	\begin{figure}[htbp]
		\centering
		\includegraphics[width=0.5\linewidth]{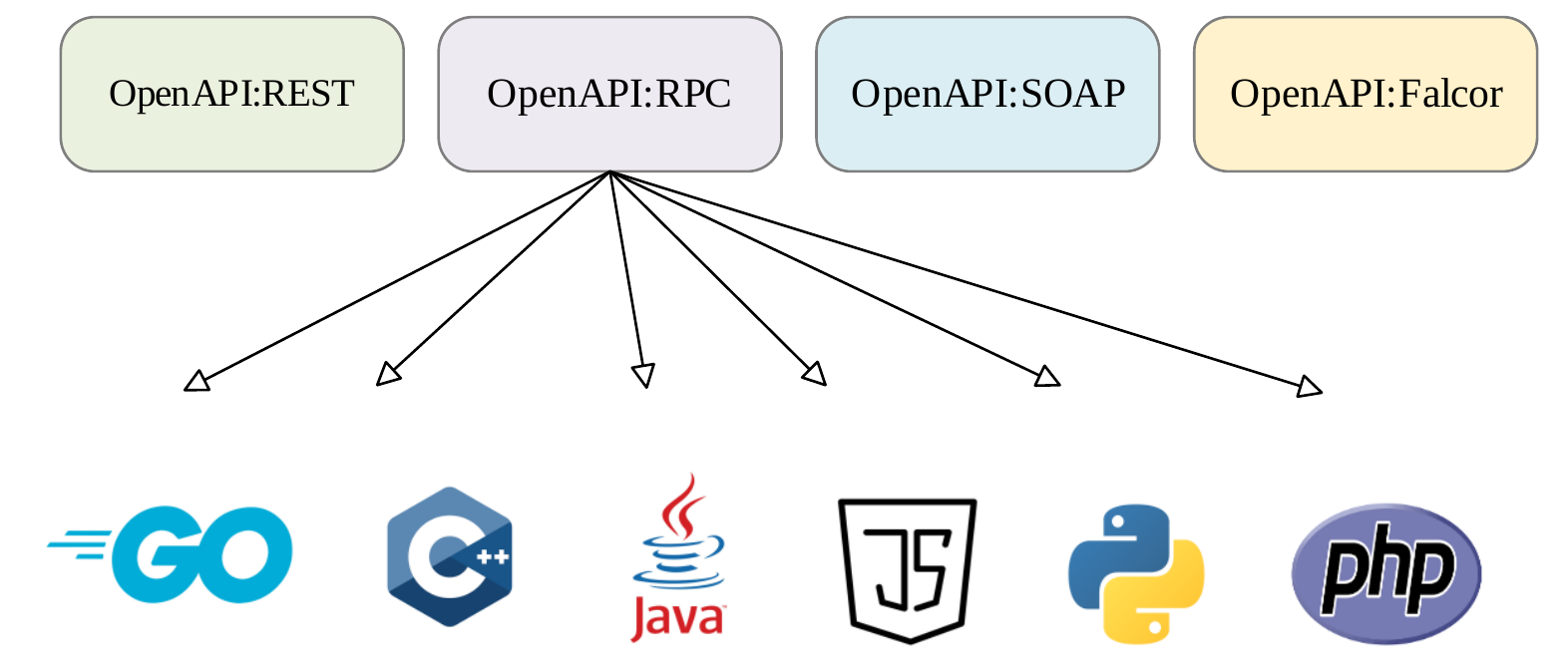}
		\caption{Traditional SDK Generation}
		\label{existingSDK}
	\end{figure}
	Inspired above, we investigated TeaDSL\footnote{https://github.com/aliyun/darabonba} \footnote{https://github.com/aliyun/darabonba-openapi}, a multi-language SDK solution for all OpenAPI gateways. Different existing SDK generation principles, our method decouple SDK from gateways(See Figure~\ref{TeaDSL}).
	
	In summary, the contributions are outlined as follows:
	\begin{enumerate}
		\item We design an API workbench(https://next.api.alibabacloud.com) to help users learn and use APIs, and we are willing to share all the data on this workbench.
		\item We explore a user experience-first method to set massive API documents' optimization order and plan.
		\item We present a real-time, fine-grained, and continuous method for evolving API documentation.
		\item We propose TeaDSL, which can be used as an intermediate language that supports different API gateways. TeaDSL allows the generation of SDKs in different languages with unified intermediate representation. 
	\end{enumerate}

	\section{Four-quadrant Management}
	In this section, we propose a four-quadrant management method to arrange the order of API optimization. The abscissa is the document coverage rate, and the ordinate is the API trial success rate. 
	\begin{figure}[H]
		\centering
		\includegraphics[width=0.5\linewidth]{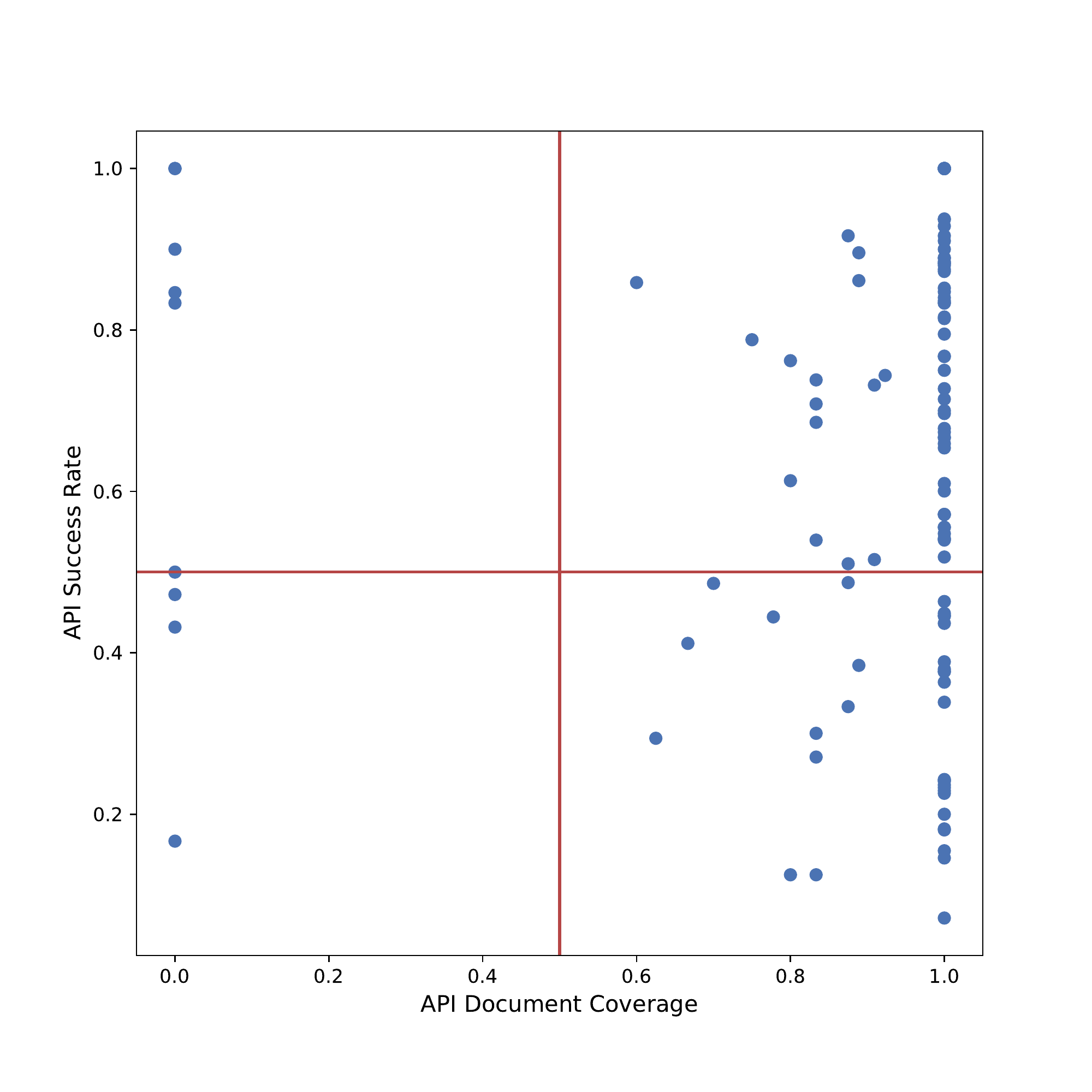}
		\caption{API Four-quadrant Management}
		\label{fig:api_opt}
	\end{figure}
	
	Each point in Figure~\ref{fig:api_opt} represents an API. The API of the third quadrant should be governed first (i.e., improve document coverage). APIs in the fourth quadrant should be manually intervened. The governance measures for these APIs may improve the quality of documents. The governance priority of APIs in the first and second quadrants can be lower.
	\section{Fine-grained  Optimization}
	
	Take the Alibaba Cloud OpenAPI developer portal as an example. Many users learn to call and debug API through the description of the document on the portal. Hence, the User's debugging record (input parameters entered by the User) represents the User's understanding of the document. If a parameter's description or example is not clear enough, the User will frequently make mistakes in the parameter's input. Thus, our core idea is:
	
	Comparing the correct parameter input and incorrect input of the same User calling the same API, the difference between the parameters is that users are prone to errors.
	
	We divide the errors into two categories:
	\begin{itemize}
		\item The parameter sequence is different between the correct and wrong calls. For instance, some parameters are missing from the wrong parameter sequence.
		\item The parameter value is different between the correct call and the wrong call.
	\end{itemize}
	
	We employ a running example to illustrate our algorithm. Assuming that we have a valid request(See \textit{\#Correct}) and a wrong request(See \textit{\#Wrong}) from the same User, we first split all request parameters to the finest granularity. For example, we translate "TemplateParam":"\{"code": "123123"\}" to "TemplateParam.code": "123123". After that, we will compare the difference between the correct request and the wrong request. Crucially, we use "-" to indicate that the reason for the request error is the lack of parameters and "+" to indicate that there are extra parameters.
	
	\textbf{A Running Example}
	\begin{lstlisting}[language = C++, numbers=left, 
		numberstyle=\tiny,keywordstyle=\color{blue!70},
		commentstyle=\color{red!50!green!50!blue!50},frame=shadowbox,
		rulesepcolor=\color{red!20!green!20!blue!20},basicstyle=\ttfamily]
#Correct
{
 "PhoneNumbers":"177xxxx9887",
 "SignName":Peking University Hospital,
 "TemplateCode":"SMS_180240289",
 "TemplateParam":"{"code":"123123"}"
}
#Wrong
{
 "PhoneNumbers":"157xxxx0621",
 "SignName": Peking University,
 "TemplateCode": successful endoscopy
}
	\end{lstlisting}
	
	After the algorithm is processed, the output is,
	\begin{lstlisting}[language = C++, numbers=left, 
		numberstyle=\tiny,keywordstyle=\color{blue!70},
		commentstyle=\color{red!50!green!50!blue!50},frame=shadowbox,
		rulesepcolor=\color{red!20!green!20!blue!20},basicstyle=\ttfamily]
[
 '-TemplateParam', #TemplateParam is missing
 '-TemplateParam.code', 
 'PhoneNumbers', #PhoneNumbers are different
 'SignName', 
 'TemplateCode' 
]
	\end{lstlisting}

	\section{TeaDSL}
	Swagger is currently a general solution for generating multilingual SDKs. However, Swagger only serves RESTful style APIs, making it impossible for many other styles to use Swagger. For example, Alibaba Cloud has many different styles of API gateways, different signature algorithms, and serialization formats, all of which are not RESTful OpenAPIs. To overcome this problem, we explore the crucial challenges of OpenAPIs and introduce a tailored solution, TeaDSL, a multi-language SDK solution for all OpenAPI gateways. TeaDSL is a domain-specific language expressing OpenAPI gateways, generating SDKs, code samples, and test cases. 
	
\begin{figure}[H]
		\centering
		\includegraphics[width=0.4\linewidth]{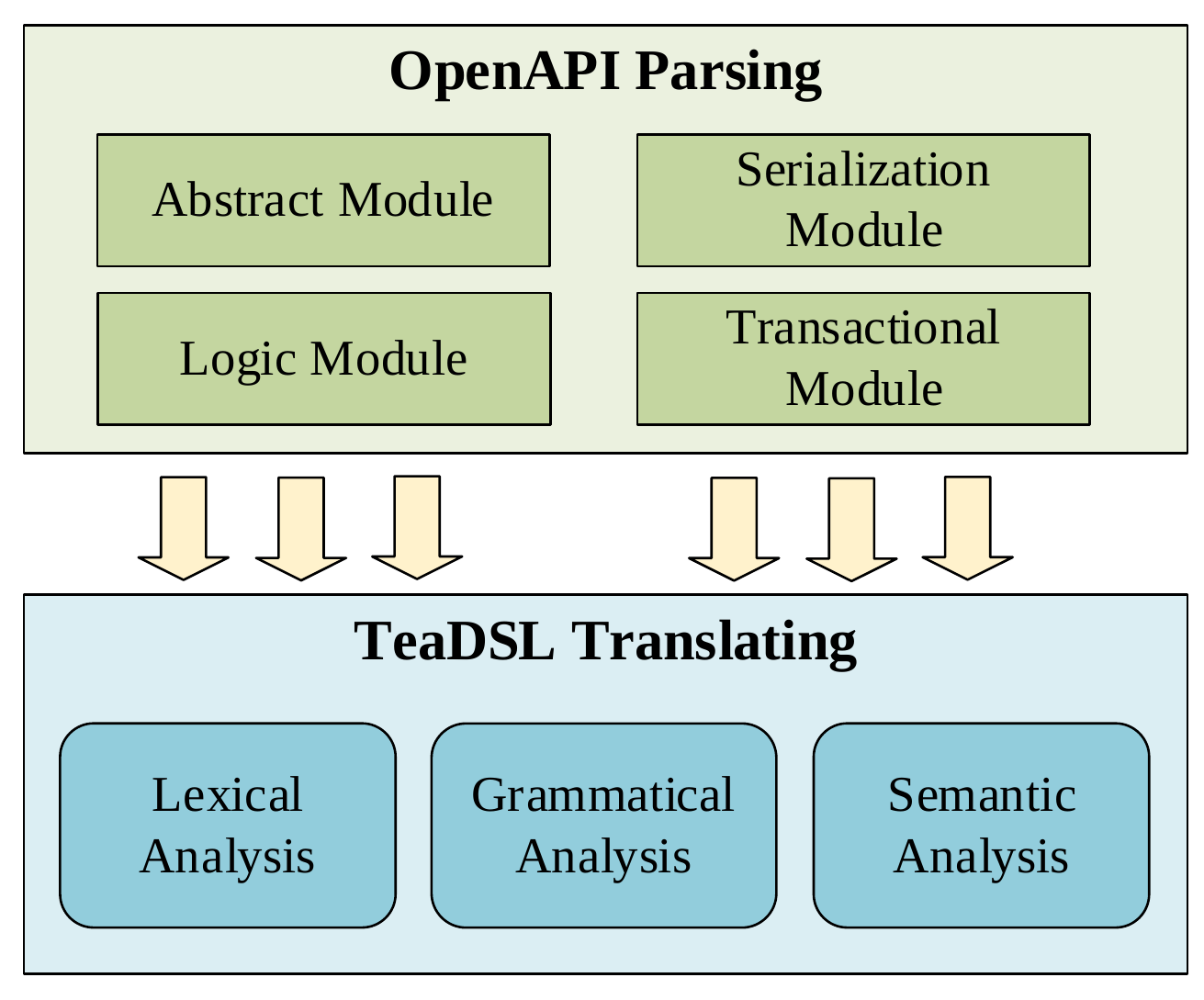}
		\caption{TeaDSL Architecture}
		\label{fig:arc}
	\end{figure}

	As shown in Figure~\ref{fig:arc}, there are two stages in TeaDSL: \textbf{OpenAPI Parsing} and \textbf{TeaDSL Translating}. TeaDSL, in the first stage, is to extract all semantics of the OpenAPI gateway description(e.g., protocol, port, request, and response) to build TeaDSL.
	Then, TeaDSL will be translated into multiple programming language SDKs.
	
	\subsection{OpenAPI Parsing}
	
	\subsubsection{Abstract Module}
	We first propose a high-level abstract approach to model all OpenAPI gateways. We investigated that all the OpenAPIs provide services based on the HTTP protocol stack. Thus, we model OpenAPI gateways as follows:
	\begin{lstlisting}[language = C++, numbers=left, 
		numberstyle=\tiny,keywordstyle=\color{blue!70},
		commentstyle=\color{red!50!green!50!blue!50},frame=shadowbox,
		rulesepcolor=\color{red!20!green!20!blue!20},basicstyle=\ttfamily]
{
 protocol: string, //HTTP or HTTPS
 port: number, //TCP port
 host: string,  //domain
 request: {
  method: string, //HTTP method
  pathname: string, //path name
  query: map[string]string, //query string
  headers: map[string]string, //headers
  body: readable  //request body
 },
 response: {
  statusCode: number, //HTTP method
  statusMessage: string, //path name
  headers: map[string]string,  //headers
  body: readable  //response body
 },
}
	\end{lstlisting}
	
	TeaDSL obeys HTTP-oriented programming principles.
	\subsubsection{Serialization Module}
	In this subsection, we introduce how to support different serialization formats. In TeaDSL, we introduce the concept of \textit{model} to describe data structure. For example,
	\begin{lstlisting}[language = C++, numbers=left, 
		numberstyle=\tiny,keywordstyle=\color{blue!70},
		commentstyle=\color{red!50!green!50!blue!50},frame=shadowbox,
		rulesepcolor=\color{red!20!green!20!blue!20},basicstyle=\ttfamily]
model User {
 username: string(pattern='[a-zA-Z1-9]'),
 age: number(pattern='\\d+', min=18,max=99)
}
	\end{lstlisting}
	
	Based on the concept of \textit{model}, we can declare a serialization method as follows:
	\begin{lstlisting}[language = C++, numbers=left, 
		numberstyle=\tiny,keywordstyle=\color{blue!70},
		commentstyle=\color{red!50!green!50!blue!50},frame=shadowbox,
		rulesepcolor=\color{red!20!green!20!blue!20},basicstyle=\ttfamily]
type @toJSONString = (User): string
	\end{lstlisting}
	
	Notice that we declare the behavior, not the specific implementation. Individual performance is guaranteed before the final operation.
	
	Finally, we use TeaDSL to package OpenAPI request/response into a method similar to programming code, for instance:
	\begin{lstlisting}[language = C++, numbers=left, 
		numberstyle=\tiny,keywordstyle=\color{blue!70},
		commentstyle=\color{red!50!green!50!blue!50},frame=shadowbox,
		rulesepcolor=\color{red!20!green!20!blue!20},basicstyle=\ttfamily]
import Util;
api getUser(username: string): User {
 __request.method = 'GET';
 __request.pathname = `/users/${username}`;
 __request.headers = {
 	host = 'hostname',
 };
} returns {
 var body = Util.readAsJSON(__response.body);
 return body;
}
	\end{lstlisting}
	
	\subsubsection{Logic Module}
	To make TeaDSL logical, we directly introduce process control(e.g., if/else if) in the programming language to express complex logic.
	\subsubsection{Transactional Module}
	Transactional expression, unlike processing request and response with sequence. Thus, we can express it through configuration.
	\subsection{TeaDSL Translating}
	This subsection emphasizes the problem of translating TeaDSL into multiple programming language SDKs. Unlike existing template-based methods to generate code, TeaDSL is a DSL code with grammar, morphology, and semantic rules. 
	
	As shown in Figure\ref{fig:arc}, our TeaDSL translating stage comprises three modules, i.e., Lexical Analysis, Grammatical analysis, and Semantic Analysis.
	
	\section{Related Work}
	Several papers in the existing literature have focused on identifying the characteristics that make an API usable based on case studies. Robillard studied API usability by surveying 83 software developers at Microsoft \cite{robillard2011field}. They found that $78 \%$ of the survey participants read API documentation to learn the APIs, $55 \%$ used code examples, $34 \%$ experimented with the APIs, $30 \%$ read articles, and $29 \%$ asked colleagues. Robillard et al. found that the most severe API learning obstacles are related to API documentation. They suggested the following requirements as must-haves for API documentation: include good examples, be complete, support many example usage scenarios, be conveniently organized, and include relevant design elements. Myers et al. also recognized documentation as a critical component for API usability and suggested using examples in the documentation to answer API-related questions \cite{myers2016improving}. Zibran et al. found that $27.3 \%$ of the reported bugs are API documentation bugs studying repositories for 562 API usability-related bugs from five different projects\cite{zibran2011useful}. Scheller et al. provided a framework for measuring API usability based on the number and types of different objects and methods that the API provides\cite{scheller2015automated}.
	
	Swagger Codegen can simplify the build process by generating server stubs and client SDKs for any API defined with the OpenAPI (formerly known as Swagger) specification(The OpenAPI Specification (OAS) defines a standard, language-agnostic interface to RESTful APIs).
	
	\section{Evaluation and Discussion} 
	
	\begin{enumerate}
		\item \textbf{Four-quadrant Management:} We use the correlation between API completeness (API document coverage) and API trial success rate to construct a four-quadrant management plan for API governance and optimization. Related engineers in the actual industrial scene have widely praised the solution.
		
		\item \textbf{Continuous Documentation Optimization:}  We employ a simple case to demonstrate our fine-grained document optimization solution, for example., API \textit{DescribeInstances} is used to query the detailed information of one or more ECS instances. The success rate of this API call is only 52.6\%.
		\begin{table}[H]
			\caption{Error Analysis\label{tab:paramSeqRate}}
			\begin{tabular}{c|c|c} 
				\hline
				\textbf{Errorcode} & \textbf{Parameter} & \textbf{Rate}\\
				\hline 
				InvalidParameter & +InstanceIds & 0.21\\
				InvalidParameter & InstanceIds & 0.14  \\
				InvalidParameter & RegionId & 0.11  \\
				\hline
			\end{tabular}
		\end{table}

		As shown in Table~1,  We found that InstanceIds is an important cause of API call errors (about 35\%). In the original document, this parameter is non-required. Its description is "Instance ID. The value can be composed of multiple instance IDs to form a JSON array, up to 100 IDs are supported, and the IDs are separated by a comma (,)". Its example is: ["i-bp67acfmxazb4p****", "i-bp67acfmxazb4p****"]. Through further analysis of the correct and incorrect calls of the User, when there is only one instance id, the User's input is no longer in the form of an array, such as "i-bp67acfmxazb4p****" instead of ["i-bp67acfmxazb4p****"]. We changed the example of this parameter to ["i-bp67acfmxazb4p****"], and now the success rate of this API call is about 76\%.
		
		\item \textbf{Availability of TeaDSL :} As shown in Figure~\ref{TeaDS_uvL}, more than 40,000 users use the SDK generated by TeaDSL to call API. TeaDSL supports the generation of 7 programming languages. We guarantee the availability of SDK generated by TeaDSL>99\% (provided that the metadata is entered correctly).
		
		\item \textbf{TeaDSL vs Swagger :} When the input metadata meets the requirements, TeaDSL and Swagger can ensure the SDK's correctness. Therefore, we compare these two tools in terms of user experience, such as usability, ease of use, stability, etc. By analyzing more than 100 developer feedback, we found that the overall evaluation of TeaDSL is better than Swagger.
		\begin{figure}[htbp]
			\centering
			\caption{TeaDSL UV}
			\includegraphics[width=0.78\linewidth]{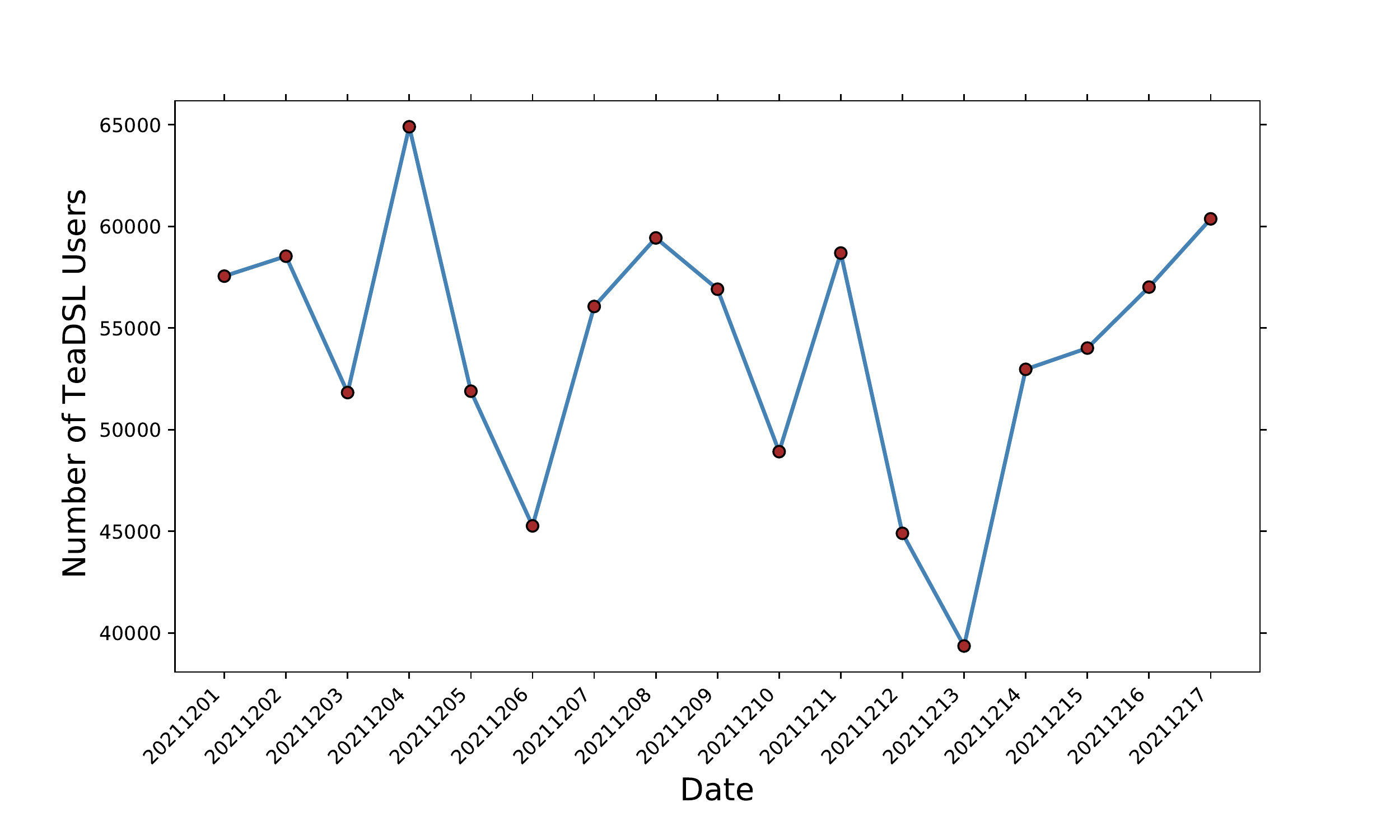}
			\label{TeaDS_uvL}
		\end{figure}
		
	\end{enumerate}
	\section*{CONCLUSIONS}
	OpenAPIs play an essential role in modern software development. With the help of APIs, developers can complete their tasks more efficiently. However, "Ease of use OpenAPIs" is an empirical challenge for end-users of the knowledge gap between API designers and API users. To mind this gap, we develop an OpenAPI workbench to help end-users learn and debug OpenAPIs. Specifically, we present a data-driven and fine-grained API documentation continuous optimization solution. We employ TeaDSL to generate multilingual SDK to enrich API documentation. The main features of TeaDSL are that it supports OpenAPIs of different styles and the generation of SDKs and code samples in multiple languages. The ultimate goal is consistency in all OpenAPI usage scenarios, such as definition, documentation, SDK generation, and CLI support. Based on the Alibaba Cloud Open Platform, we extend TeaDSL through integration with more accessible tools to build a more compatible ecosystem than Swagger.
	
	\section*{Acknowledgments}
	
	This work was supported by Alibaba Group through Alibaba Innovative Research Program.

	\bibliographystyle{ACM-Reference-Format} \nocite{1}
	\bibliography{reference}


\begin{thebibliography}{11}


\ifx \showCODEN    \undefined \def \showCODEN     #1{\unskip}     \fi
\ifx \showDOI      \undefined \def \showDOI       #1{#1}\fi
\ifx \showISBNx    \undefined \def \showISBNx     #1{\unskip}     \fi
\ifx \showISBNxiii \undefined \def \showISBNxiii  #1{\unskip}     \fi
\ifx \showISSN     \undefined \def \showISSN      #1{\unskip}     \fi
\ifx \showLCCN     \undefined \def \showLCCN      #1{\unskip}     \fi
\ifx \shownote     \undefined \def \shownote      #1{#1}          \fi
\ifx \showarticletitle \undefined \def \showarticletitle #1{#1}   \fi
\ifx \showURL      \undefined \def \showURL       {\relax}        \fi
\providecommand\bibfield[2]{#2}
\providecommand\bibinfo[2]{#2}
\providecommand\natexlab[1]{#1}
\providecommand\showeprint[2][]{arXiv:#2}

\bibitem[\protect\citeauthoryear{Gu, Zhang, Zhang, and Kim}{Gu
  et~al\mbox{.}}{2016}]%
        {APIEffecient}
\bibfield{author}{\bibinfo{person}{Xiaodong Gu}, \bibinfo{person}{Hongyu
  Zhang}, \bibinfo{person}{Dongmei Zhang}, {and} \bibinfo{person}{Sunghun
  Kim}.} \bibinfo{year}{2016}\natexlab{}.
\newblock \showarticletitle{Deep {API} learning}. In
  \bibinfo{booktitle}{\emph{Proceedings of the 24th {ACM} {SIGSOFT}
  International Symposium on Foundations of Software Engineering, {FSE} 2016,
  Seattle, WA, USA, November 13-18, 2016}},
  \bibfield{editor}{\bibinfo{person}{Thomas Zimmermann}, \bibinfo{person}{Jane
  Cleland{-}Huang}, {and} \bibinfo{person}{Zhendong Su}} (Eds.).
  \bibinfo{publisher}{{ACM}}, \bibinfo{pages}{631--642}.
\newblock
\urldef\tempurl%
\url{https://doi.org/10.1145/2950290.2950334}
\showDOI{\tempurl}


\bibitem[\protect\citeauthoryear{Huang, Xia, Xing, Lo, and Wang}{Huang
  et~al\mbox{.}}{2018}]%
        {API}
\bibfield{author}{\bibinfo{person}{Qiao Huang}, \bibinfo{person}{Xin Xia},
  \bibinfo{person}{Zhenchang Xing}, \bibinfo{person}{David Lo}, {and}
  \bibinfo{person}{Xinyu Wang}.} \bibinfo{year}{2018}\natexlab{}.
\newblock \showarticletitle{API method recommendation without worrying about
  the task-API knowledge gap}. In \bibinfo{booktitle}{\emph{2018 33rd IEEE/ACM
  International Conference on Automated Software Engineering (ASE)}}. IEEE,
  \bibinfo{pages}{293--304}.
\newblock


\bibitem[\protect\citeauthoryear{Lamothe}{Lamothe}{2020}]%
        {gapbetuseranddesigner}
\bibfield{author}{\bibinfo{person}{Maxime Lamothe}.}
  \bibinfo{year}{2020}\natexlab{}.
\newblock \showarticletitle{Bridging the divide between {API} users and {API}
  developers by mining public code repositories}. In
  \bibinfo{booktitle}{\emph{{ICSE} '20: 42nd International Conference on
  Software Engineering, Companion Volume, Seoul, South Korea, 27 June - 19
  July, 2020}}, \bibfield{editor}{\bibinfo{person}{Gregg Rothermel} {and}
  \bibinfo{person}{Doo{-}Hwan Bae}} (Eds.). \bibinfo{publisher}{{ACM}},
  \bibinfo{pages}{178--181}.
\newblock
\urldef\tempurl%
\url{https://doi.org/10.1145/3377812.3382124}
\showDOI{\tempurl}


\bibitem[\protect\citeauthoryear{Myers and Stylos}{Myers and Stylos}{2016}]%
        {myers2016improving}
\bibfield{author}{\bibinfo{person}{Brad~A Myers} {and} \bibinfo{person}{Jeffrey
  Stylos}.} \bibinfo{year}{2016}\natexlab{}.
\newblock \showarticletitle{Improving API usability}.
\newblock \bibinfo{journal}{\emph{Commun. ACM}} \bibinfo{volume}{59},
  \bibinfo{number}{6} (\bibinfo{year}{2016}), \bibinfo{pages}{62--69}.
\newblock


\bibitem[\protect\citeauthoryear{Nguyen, Nguyen, Phan, and Nguyen}{Nguyen
  et~al\mbox{.}}{2017}]%
        {apiuse10}
\bibfield{author}{\bibinfo{person}{Trong~Duc Nguyen}, \bibinfo{person}{Anh~Tuan
  Nguyen}, \bibinfo{person}{Hung~Dang Phan}, {and} \bibinfo{person}{Tien~N.
  Nguyen}.} \bibinfo{year}{2017}\natexlab{}.
\newblock \showarticletitle{Exploring {API} embedding for {API} usages and
  applications}. In \bibinfo{booktitle}{\emph{Proceedings of the 39th
  International Conference on Software Engineering, {ICSE} 2017, Buenos Aires,
  Argentina, May 20-28, 2017}},
  \bibfield{editor}{\bibinfo{person}{Sebasti{\'{a}}n Uchitel},
  \bibinfo{person}{Alessandro Orso}, {and} \bibinfo{person}{Martin~P.
  Robillard}} (Eds.). \bibinfo{publisher}{{IEEE} / {ACM}},
  \bibinfo{pages}{438--449}.
\newblock
\urldef\tempurl%
\url{https://doi.org/10.1109/ICSE.2017.47}
\showDOI{\tempurl}


\bibitem[\protect\citeauthoryear{Robillard and DeLine}{Robillard and
  DeLine}{2011}]%
        {robillard2011field}
\bibfield{author}{\bibinfo{person}{Martin~P Robillard} {and}
  \bibinfo{person}{Robert DeLine}.} \bibinfo{year}{2011}\natexlab{}.
\newblock \showarticletitle{A field study of API learning obstacles}.
\newblock \bibinfo{journal}{\emph{Empirical Software Engineering}}
  \bibinfo{volume}{16}, \bibinfo{number}{6} (\bibinfo{year}{2011}),
  \bibinfo{pages}{703--732}.
\newblock


\bibitem[\protect\citeauthoryear{Scheller and K{\"u}hn}{Scheller and
  K{\"u}hn}{2015}]%
        {scheller2015automated}
\bibfield{author}{\bibinfo{person}{Thomas Scheller} {and} \bibinfo{person}{Eva
  K{\"u}hn}.} \bibinfo{year}{2015}\natexlab{}.
\newblock \showarticletitle{Automated measurement of API usability: The API
  concepts framework}.
\newblock \bibinfo{journal}{\emph{Information and Software Technology}}
  \bibinfo{volume}{61} (\bibinfo{year}{2015}), \bibinfo{pages}{145--162}.
\newblock


\bibitem[\protect\citeauthoryear{Shen, Wu, Zou, Zhu, and Xie}{Shen
  et~al\mbox{.}}{2020}]%
        {apiuse12}
\bibfield{author}{\bibinfo{person}{Qi Shen}, \bibinfo{person}{Shijun Wu},
  \bibinfo{person}{Yanzhen Zou}, \bibinfo{person}{Zixiao Zhu}, {and}
  \bibinfo{person}{Bing Xie}.} \bibinfo{year}{2020}\natexlab{}.
\newblock \showarticletitle{From {API} to {NLI:} {A} new interface for library
  reuse}.
\newblock \bibinfo{journal}{\emph{J. Syst. Softw.}}  \bibinfo{volume}{169}
  (\bibinfo{year}{2020}), \bibinfo{pages}{110728}.
\newblock
\urldef\tempurl%
\url{https://doi.org/10.1016/j.jss.2020.110728}
\showDOI{\tempurl}


\bibitem[\protect\citeauthoryear{Wen, Liu, Wu, Xie, Cheung, and Su}{Wen
  et~al\mbox{.}}{2019}]%
        {apiuse11}
\bibfield{author}{\bibinfo{person}{Ming Wen}, \bibinfo{person}{Yepang Liu},
  \bibinfo{person}{Rongxin Wu}, \bibinfo{person}{Xuan Xie},
  \bibinfo{person}{Shing{-}Chi Cheung}, {and} \bibinfo{person}{Zhendong Su}.}
  \bibinfo{year}{2019}\natexlab{}.
\newblock \showarticletitle{Exposing library {API} misuses via mutation
  analysis}. In \bibinfo{booktitle}{\emph{Proceedings of the 41st International
  Conference on Software Engineering, {ICSE} 2019, Montreal, QC, Canada, May
  25-31, 2019}}, \bibfield{editor}{\bibinfo{person}{Joanne~M. Atlee},
  \bibinfo{person}{Tevfik Bultan}, {and} \bibinfo{person}{Jon Whittle}} (Eds.).
  \bibinfo{publisher}{{IEEE} / {ACM}}, \bibinfo{pages}{866--877}.
\newblock
\urldef\tempurl%
\url{https://doi.org/10.1109/ICSE.2019.00093}
\showDOI{\tempurl}


\bibitem[\protect\citeauthoryear{Zhong, Meng, Li, and Jia}{Zhong
  et~al\mbox{.}}{2020}]%
        {apiuse1}
\bibfield{author}{\bibinfo{person}{Hao Zhong}, \bibinfo{person}{Na Meng},
  \bibinfo{person}{Zexuan Li}, {and} \bibinfo{person}{Li Jia}.}
  \bibinfo{year}{2020}\natexlab{}.
\newblock \showarticletitle{An empirical study on {API} parameter rules}. In
  \bibinfo{booktitle}{\emph{{ICSE} '20: 42nd International Conference on
  Software Engineering, Seoul, South Korea, 27 June - 19 July, 2020}},
  \bibfield{editor}{\bibinfo{person}{Gregg Rothermel} {and}
  \bibinfo{person}{Doo{-}Hwan Bae}} (Eds.). \bibinfo{publisher}{{ACM}},
  \bibinfo{pages}{899--911}.
\newblock
\urldef\tempurl%
\url{https://doi.org/10.1145/3377811.3380922}
\showDOI{\tempurl}


\bibitem[\protect\citeauthoryear{Zibran, Eishita, and Roy}{Zibran
  et~al\mbox{.}}{2011}]%
        {zibran2011useful}
\bibfield{author}{\bibinfo{person}{Minhaz~F Zibran}, \bibinfo{person}{Farjana~Z
  Eishita}, {and} \bibinfo{person}{Chanchal~K Roy}.}
  \bibinfo{year}{2011}\natexlab{}.
\newblock \showarticletitle{Useful, but usable? factors affecting the usability
  of APIs}. In \bibinfo{booktitle}{\emph{2011 18th Working Conference on
  Reverse Engineering}}. IEEE, \bibinfo{pages}{151--155}.
\newblock


\end{thebibliography}

\end{document}